\documentstyle[epsfig]{elsart}
\begin{document}
\begin{frontmatter}
\title{
Monte Carlo simulation of electromigration phenomena in metallic lines}
\author{ C. Pennetta, L. Reggiani and  E. Alfinito}
\address{
INFM - National Nanotechnology Laboratory, Dipartimento di Ingegneria \\
dell'Innovazione, Universit\`a di Lecce, 
Via Arnesano, I-73100 Lecce, Italy}
%
%
\begin{abstract}
The electromigration (EM) of metallic lines is studied in terms of competition
between two percolative processes taking place in a random resistor network. 
The effects associated with the transport of mass and with the consequent 
growth of the internal stress are accounted for by stochastic generation 
and recovery of voids, driven by the external current. 
Monte Carlo simulations enable us to investigate within a unified theoretical 
framework a variety of relevant aspects of EM degradation as: damage patterns,
the distribution of the time to failures, early stage resistance change 
in the presence of compositional effects, Black's law, geometrical effects, 
etc. A general agreement is found with the experimental findings.
\end{abstract}            
\thanks {This work has been performed within the STATE project
of the INFM. Partial support from the MADESS II of the Italian 
CNR as well as discussions with 
Drs. I. De Munari, F. Fantini, A. Scorzoni and G. Trefan
are gratefully acknowledged.}
\end{frontmatter}
	\section{Introduction}
In recent times, the study of breakdown phenomena in electrical devices 
has received a lot of attention, mainly for the increasing level of 
miniaturization which makes failure problems more and more 
crucial \cite{ohring}. 
Failure occurs in many cases by the degradation of metallic interconnects 
(thin films) which, because of electromigration (EM), 
lose their conducting properties. EM arises from the transport 
of matter at the atomic level driven by a high current 
density \cite{ohring,scorzoni92,fantini98}. 
More precisely, the damage, consisting in the formation and growth 
of voids and hillocks in different regions of the film, is due to a 
non-steady atomic trasport associated with a nonvanishing divergence of the 
atomic flux.
A fundamental ingredient in the description of the EM damage of the 
interconnects in electronic devices is represented by the granular 
structure of the materials employed, Al, Cu, Al alloys, etc. 
In fact, it has been observed \cite{ohring,scorzoni92,fantini98} that the 
atomic transport through the grain boundaries (transport channels) far 
exceeds that through the grain bulks. Therefore, it is generally possible 
to neglect mass transport everywhere except within these channels and to 
describe the film as an interconnected grain boundary network \cite{ohring}.
Futhermore, a high degree of disorder is usually present in alloy films due 
to compositional effects (CE) and to thermal 
gradients \cite{ohring,scorzoni92,fantini98}. Finally, it must be 
underlined the r\^ole of mechanical stress. Indeed, the depletion and 
accumulation of mass in different regions of the film, implies also the 
growth of mechanical stress, which contrasts the atomic flux due to EM 
\cite{ohring,scorzoni92,blech}. 
On the basis of the above description, we have developed a model which 
describes the EM damage of interconnects in terms of two competing biased 
percolations taking place in a Random Resistor Network (RRN) \cite{stauffer}. 
Monte Carlo simulations allow us to explore a variety of relevant aspects 
of EM degradation as: damage patterns, the distribution of the time to 
failures, early stage resistance change in the presence of compositional 
effects, Black's law, geometrical effects, etc. The paper is organized as 
follows. Section 2 describes the model used and the numerical procedure of 
simulations. Section 3 reports the results and compares theory with well 
accepted phenomenological behaviors. Major conclusions are drawn in Sec. 4.
	\section{Model}
A metallic line of length $L$ and width $w$ is described as a
two-dimensional rectangular-lattice network of regular resistors,
$r_{reg}$. The network lies on an insulating substrate at temperature
$T_0$ acting as a thermal reservoir.  The total number of elementary
resistors in the network is given by: $N_{tot}= 2N_{L}N_{w} + N_{L} -
N_{w}$, where $ N_{L}$ and $N_{w}$ determine the length and the width
of the network. To reduce the computational efforts, long lines are
simulated by assuming that only a portion of the line is responsible
of the resistance variations. The network represents this portion 
and its length is taken $1/F$ times smaller than the line length. 
Thus, the relative resistance variation of the whole line is obtained by 
multiplying the relative resistance variations of the network by the 
factor $F$. The resistance of the $n$-th resistor depends 
on temperature according with: 
\begin{equation} 
r_{reg,n} (T_n)=r_{ref}[ 1 + \alpha (T_n - T_{ref})]
\label{eq:temp_dep}
\end{equation}
where $\alpha$ is the temperature coefficient of resistance (TCR), 
$T_n$ is the local temperature, $T_{ref}$ and $r_{ref}$ are the reference 
values for the TCR. The network is contacted at the left and right hand sides 
to perfectly conducting bars through which a constant stress current $I$ is 
applied. The Joule heating induced by $I$ is taken into account by defining 
$T_n$ as \cite{pen_prl00}:
\begin{equation}
T_n=T_{0} + A \Bigl[ r_n i_n^{2} + {B \over N_{neig}} \sum_{m=1}^{N_{neig}}
 \Bigl( r_{m,n} i_{m,n}^2   - r_n i_n^2 \Bigr) \Bigr],  \label{eq:temp}
\end{equation}
where $A$ is the thermal resistance of the single resistor and $N_{neig}$
the number of first neighbours of the $n$-th resistor. The value $B=3/4$ is 
chosen to provide a uniform heating for the perfect network \cite{pen_prl00}.
By taking this expression we are assuming an istantaneous thermalization. 
In the presence of a current stress, the EM damage is simulated by allowing 
the breaking of regular resistors, i.e. by replacing 
$r_{reg} \rightarrow  r_{OP}$, $r_{OP} = 10^9 \ r_{ref}$. We remark that $r_{OP}$ 
represents an open circuit (OP) associated with the formation of voids 
inside the line. Simultaneously, a recovery process which allows the healing of the broken resistors, is also considered.
The complete failure of the line is thus associated with the existence of at
least one continuous path of OP between the upper and lower sides of the
network (percolation threshold) \cite{stauffer}.
The probability $W_{OP}$ for the $n$-th resistor to become OP is taken as: 
\begin{equation}
W_{OP}=\exp( -{E_{OP} / k_BT_n} ) 
\label{eq:defcrea}
\end{equation}
where $E_{OP}$ is a characteristic activation energy and $k_B$
the Boltzmann constant. A similar expression is taken for the recovery 
process probability  $W_{RE}$, with $E_{OP}$ replaced by $E_{RE}$.
Accordingly, the elimination of resistors represents the mass transport due 
to an electron wind, and  the recovery of resistors, the back-up flow of atoms 
due to the mechanical stress. The two activation energies control the 
competition between these two opposite processes.
	\subsection{Compositional effects (CE)}
In Al alloys (Al-Cu, Al-Si) CE are observed at the early stages of EM test of
metallic lines. Here, we limit ourself to study the CE due to: (i) a sample 
heating, (ii) an intrinsic defectiveness of the manifactured samples.
To simulate the CE induced by heating, we take into account the
actual thermal treatment undergone by the lines just before the EM test.
Here, we consider one specific EM experiment which is 
performed with standard techniques on Al-0.5\%Cu lines.
For these lines, the last thermal treatment occurred during 
fabrication was a high temperature annealing followed by a rapid cooling.
This treatment left a non-equilibrium concentration of Cu dissolved into the 
Al matrix. Therefore, in the early stage of the EM test, 
the heating associated with the stress conditions allows a relaxation toward 
the equilibrium concentration of the Cu atoms and gives rise to the formation 
of CuAl$_2$  precipitates. The variation of the line composition is 
simulated by changing, with probability $W_{ri}$, the resistance of the 
$n$-th resistor from $r_{reg}$ to $r_{imp}$ (impurity resistors), where 
we take:
\begin{equation} 
W_{ri}={\rm exp}(-{E_{ri} / k_BT_n})
\label{eq:W_ri}
\end{equation}
being $E_{ri}$ an activation energy characterizing this process. 
The mechanism of Cu dissolution into the Al matrix, antagonist to the 
precipitation, is simultaneously considered by taking a similar expression 
for the probability of the reverse process $r_{imp} \rightarrow r_{reg}$, 
which is determined by an activation energy $E_{ir}$. 
As a consequence of the two competing processes, a steady state concentration 
of $r_{imp}$ is achieved on a time scale much shorter than that of EM.
The intrinsic defectiveness of the samples manifests itself by the existence 
of samples of the same material and geometry but with drastically different 
failure times. To simulate this second case of CE, we have considered the 
evolution of networks with different initial concentration $p_{ini}$ of OP 
resistors. Accordingly, we have assumed $p_{ini}$ randomly distributed, and 
chosen values of $r_0$ in such a way to keep nearly the same 
resistance value of similar undefected samples.
	\subsection{Numerical procedure}
	\begin{figure}[t]
	\label{fig:1}
	\epsfclipon
	\centerline{\epsfig{file=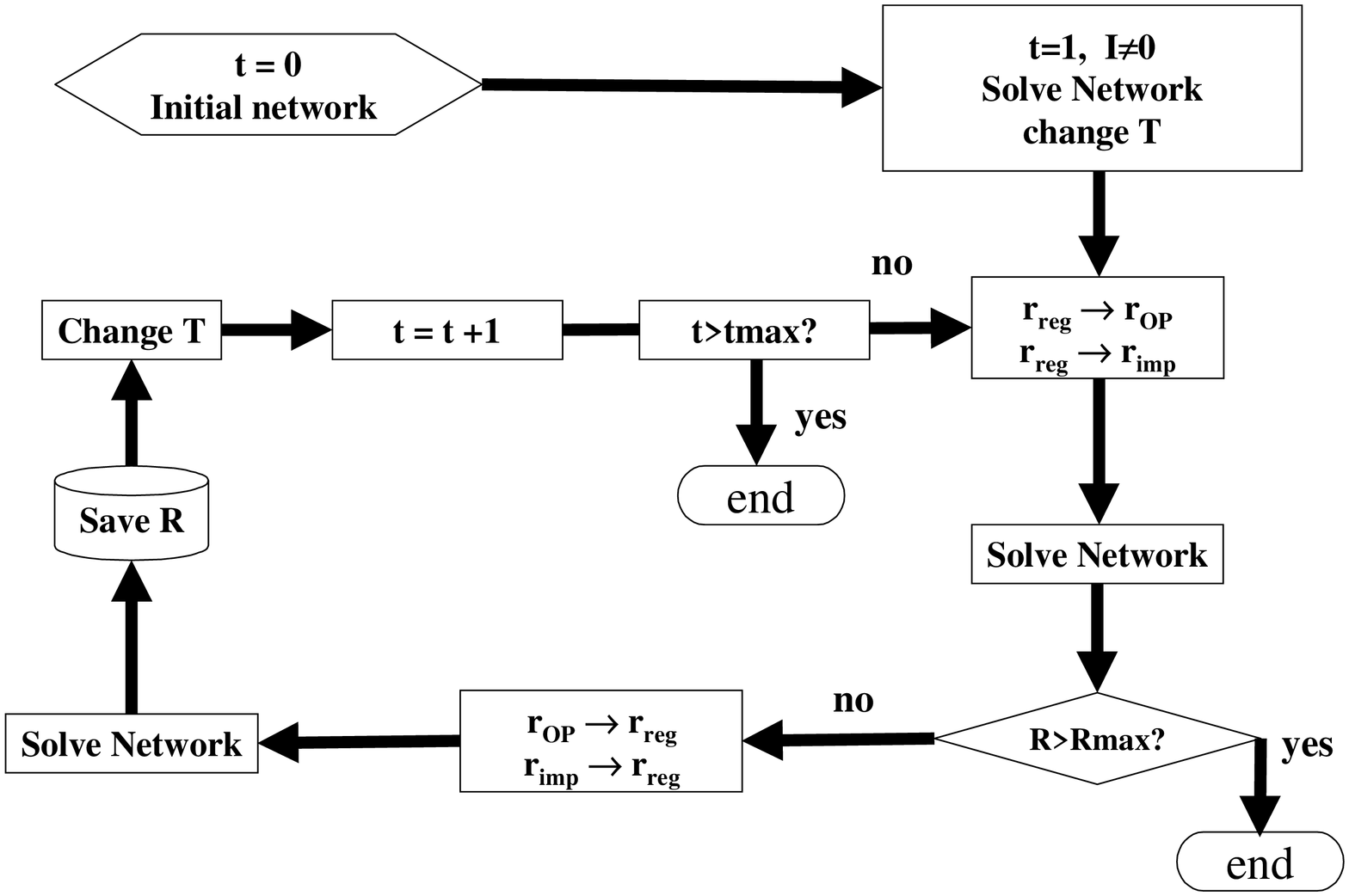,width=0.5
	\linewidth,height=0.75\linewidth,angle=-90} \vspace{.2cm}}
	\caption{\small Flowchart of the Monte Carlo simulator.}
	\end{figure}

The network evolution is obtained by Monte Carlo simulations which are 
carried out according with the flowchart shown in Fig. 1.
(i) Starting from the initial network, we calculate $i_n$ and the network 
resistance $R$ by solving Kirchhoff's loop equations. 
Moreover, we calculate $T_n$ by using Eq. (2).
(ii) OP and $r_{imp}$ are generated with the corresponding probabilities 
$W_{OP}$ and $W_{ri}$, while the remaining $r_{reg}$ are changed according 
to $T_n$. Then, $i_n$ and $T_n$ are recalculated. 
(iii) OP and $r_{imp}$ are recovered. 
(iv) $i_n$, $T_n$ and $R$ are recalculated. 
This procedure is iterated from (ii), thus the loop (ii)-(iv) 
corresponds to an iteration step, which is associated with a unit time
step on an arbitrary time scale to be calibrated by comparison 
with experiments. The iteration proceeds until, depending on the values of 
the model parameters, the following two possibilities are achieved: 
irreversible failure or steady-state evolution. The irreversible failure 
can be established by a criterion for the increase of resistance over its 
initial  value as suggested by experiments (typically of 20 \%).
	\section{Results}
	\begin{figure}[t]
	\label{fig:2}
	\epsfclipon
	\centerline{\epsfig{file=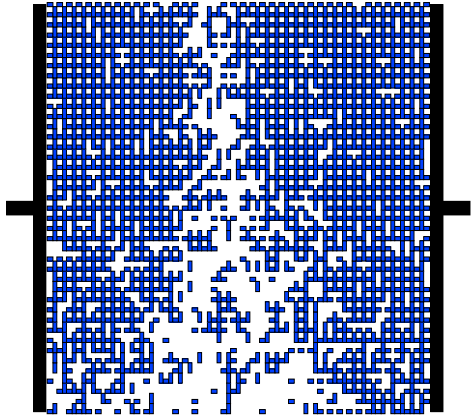,width=0.5
          \linewidth,height=0.46\linewidth}}
	\caption{\protect\small 
	Simulated damage pattern at the complete failure of a 
	network $40\times 40$ by using the parameters specified in the text.}
	\end{figure}
To check the ability of the model to describe EM experiments, simulations are
performed on Al-0.5\%Cu lines \cite{pennetta01}. 
A significant aspect related to the electrical failure is the damage
pattern, which evidences the correlations among different failing regions 
just before the complete failure.
Figure 2 reports a typical damaged pattern obtained for a network of
sizes $40 \times 40$ with the following parameters chosen according to
the experimental conditions of Ref. \cite{pennetta01}: 
$I=1.1 \times 10^{-2} \ A$, 
$T_0 =219\ ^oC$, $T_{ref}=0 \ ^oC$, $r_{ref}=296 \ \Omega$, $\alpha = 3.6
\times 10^{-3} \ ^oC^{-1}$.  
To provide for an initial heating of the network of about 15 $^oC$, the value 
of the thermal resistance $A$ is taken as: $1.5 \times 10^6 \ ^oC/W$; 
to account for an initial defectiveness of the line we take 
$p_{ini}\ \approx \ 0.01$. Moreover, to save computational time, we choose 
F=30 and for the activation energies responsible of EM we use values smaller
than those found in experiments, as: $E_{OP}$ = $E_{RE}$ = 0.43 eV.
We have tested that, by scaling the value of $E_{OP}$ in the range
$0.2 \div 0.6$ eV, the median time to failure (MTF) changes over more
than two orders of magnitude while its relative standard deviation remains
constant within a factor of two.  Consistently with this choice, to
account for the different characteristic times of the two processes,
also the activation energies responsible of CE have been reduced: we 
take $E_{ri} =0.22 \ eV$ and $E_{ir}$=0.17 eV. Moreover, for the
impurity resistor we use $r_{imp}=350 \ \Omega$.
We note that the damage pattern in Fig. 2 well reproduces the
experimental pattern observed by SEM in metallic lines \cite{ohring,fantini98}.
	\begin{figure}[t] 
	\label{fig:3} 
	\epsfclipon
	\centerline{\epsfig{file=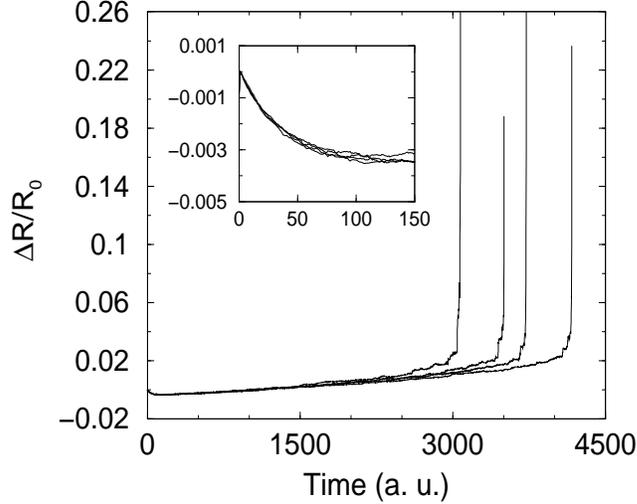,width=0.6
	\linewidth,height=0.5\linewidth}} 
	\caption{\protect\small
	Simulation of the relative variations of resistance as
	function of time for Al-0.5\%Cu lines. The insert
	shows enlarged CE in the early stage of EM.
	\vspace{0.3cm}}	
\end{figure}
	\begin{figure}[b] 
	\label{fig:4} 
	\epsfclipon 
	\centerline{\epsfig{file=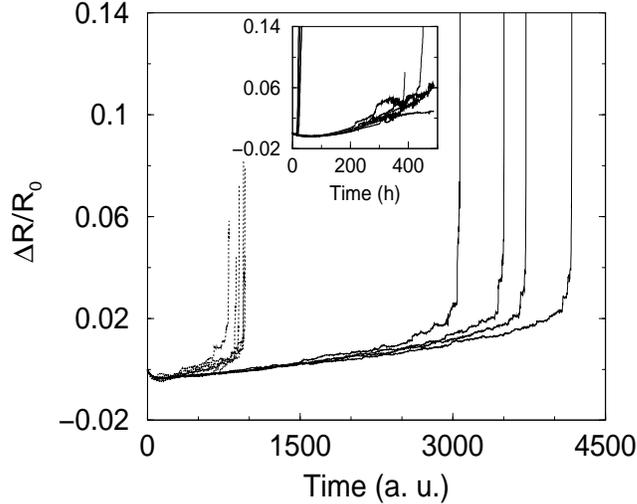,width=0.6
	\linewidth,height=0.5\linewidth}} 
	\caption{\protect\small
	Simulation of the relative variations of resistance as
	function of time corresponding to networks with different initial 
	defectiveness and comparable resistance. Continuous curves 
	represent evolutions obtained by taking $p_{ini}\approx0.1$, 
        the dotted curves are for $p_{ini} \approx0.3$.
	The insert reports a set of experimental results
	performed on samples with the same geometry and comparable 
	resistance.}		
	\end{figure}
	\begin{figure}[b] 
	\label{fig:5} 
	\epsfclipon
	\centerline{\vspace{0.3cm}\epsfig{file=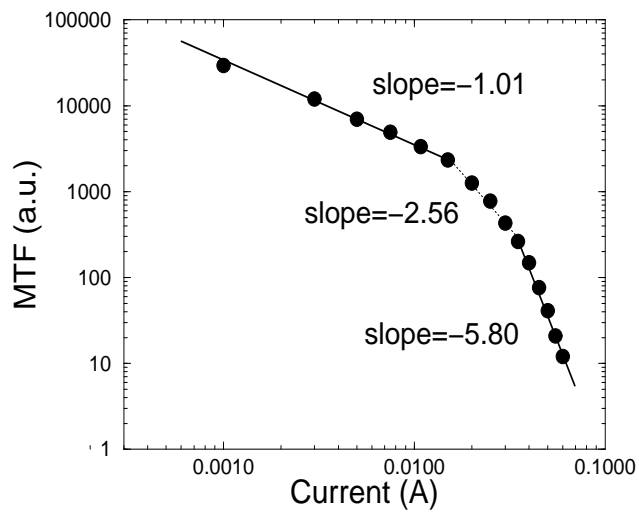,width=0.6
	\linewidth,height=0.5\linewidth}\vspace{-0.1cm}} 	
	\caption{\protect\small Median time to failure as a function
	of stressing current for a rectangular network $12 \times 400$
	with the parameter values specified in the text. The
	slopes in the relevant regions of the curve (low, moderate, and
	high current stresses) are reported in the figure.}
	\end{figure}
The results of simulation for the resistance evolutions are reported
in Fig.3.  All the parameters are the same used for Fig. 2, except
for the network sizes, that are now $N_{L}=N_{w}=70$. The insert shows the 
early stage resistance decrease due to the thermal CE. The simulations 
reproduce satisfactorily the experimental evidence of both the early stage 
and the pre-failure regime of the resistance evolution \cite{pennetta01}. 
To investigate the r\^ole of the intrinsic defectiveness of the line, we have 
simulated the evolution of the network with a different initial concentration 
of OP resistors randomly distributed. Moreover, we
have chosen $r_{ref}\ = \ 229 \ \Omega$ to have nearly the same
resistance of the networks considered in Fig. 3. We report in Fig. 4 the
relative variations of the resistance as function of time for
the two sets of simulations: continuous curves are the same of Fig. 3
($p_{ini} \approx 0.1$), the other ones corresponding to $p_{ini} \approx
0.3$. We show in the insert the results of the experiments \cite{pennetta01}
performed on samples with the same geometry and comparable resistance
but exhibiting completely different failure times. The simulations
well reproduce the experiments \cite{pennetta01}.

EM tests are performed in the so-called ``accelerate conditions'', i.e. the
current density, $J$, and the stress temperature, $T$, are much higher than 
those corresponding to normal operation conditions. This is an obliged choice
to obtain results on a time scale of days or weeks, whereas, in normal
conditions, the failure times are of the order of years. Therefore, the 
evaluation of the failure times in normal conditions, comes from the 
existence of a scaling relation connecting accelerate conditions with 
normal ones. This relation is provided by the well known Black's 
law \cite{black67} for the median time to failure (MTF): 
$$ {\rm MTF}^{-1}\,\propto\, J^{n}\ T^{-2}\ \exp\left[{\frac{-E}
{k_{B}T}}\right]$$ 
where $E$ is the activation energy for EM. For moderate stress conditions 
the value of $n$ is found between 1 and 2, while higher values are found for 
extreme stress conditions. Several models have been proposed to explain this 
law with contradictory predictions concerning the value of the exponent 
$n$ \cite{lloyd86,clement92,tammaro99}. Therefore, we have calculated the MTF 
as a function of the applied current and the results are shown in Fig. 5. The 
simulations are performed on rectangular networks with sizes $12 \times 400$
by taking the same values of $T_0$, $T_{ref}$, $\alpha$ used before and the 
results are shown in Fig. 5.
The values of the other parameters are the following: 
$r_{ref}=0.048 \ \Omega$, $r_{imp}=0.016 \ \Omega$,
$A=2.7 \times 10^8 \ ^oC/W$, $p_{ini}=0$, $F=200$,
$E_{OP}=0.41$ eV, $E_{RE}= 0.35$ eV, $E_{ri}=0.22$ eV and 
$E_{ir}=0.17$ eV. From Fig. 5 we can see that experimental behaviors are 
naturally reproduced by our simulations, i.e. $n\approx 1$ for low bias, 
and $n$ grows up to 5.8 for extreme bias values.
	\begin{figure}[b] 
	\label{fig:6} 
	\epsfclipon
	\centerline{\epsfig{file=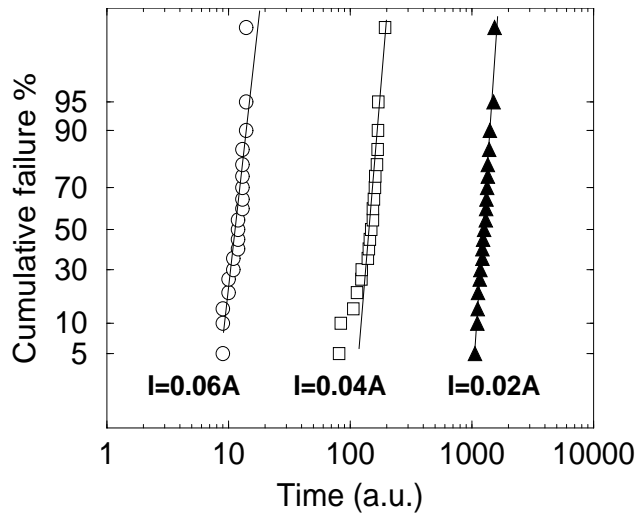,width=0.6
	\linewidth,height=0.5\linewidth}} 	
        \caption{\protect\small
	Distribution of TTFs obtained from $20$ 
	simulations of failure at different stressing currents. The 
        network sizes and the parameter values are the same of Fig. 5. 
        Lines show the fit with a log-normal distribution.
        \vspace{.2cm}} 	
	\end{figure}
It is interesting to consider the distribution of the times to failure
(TTF) of a set of identical samples in the same stress
conditions. Therefore we report in Figure 6 the distribution of the
TTF obtained by simulations, at increasing values of the stressing
current (going from the right to the left). All the parameters are the same 
of that used for Fig. 5. For each current value, 20 realizations have been 
considered. Precisely, in the figure the cumulative failure percentage is 
reported as a function of the logarithm of the TTF and a transformation has 
been applied to the ordinate axis in such a manner that a log-normal 
distribution would be a straight line.
From the figure we can conclude that the distribution of the simulated 
TTF is log-normal (within errors) and that the shape factor (the log-normal
standard deviation) is independent of the applied current. These
conclusions well agree with the experiments \cite{ohring,scorzoni92,fantini98}.
%
%
As a result of the competition between the electromigration and the
backflow of mass determined by stress gradients, the failure can occur
or not depending on the length of the line. As a consequence, it is
possible to consider this geometrical effects on the value of the MTF. 
Figure 7 shows the simulated MTF as function of the length to width ratio
of the network (with fixed width). We remark that the MTF diverges for 
small lengths, according with the experimental results \cite{scorzoni92}. 
Experiments also show that for a line of given length it exists a critical 
value of the current density below which there is no electromigration. 
Moreover, this breakdown value is inversely proportional to the length
\cite{ohring,scorzoni92,fantini98}. Our simulations well reproduce also 
this behavior as shown in Fig. 8, where the breakdown current is reported 
as a function of the width to length ratio (again at width fixed).
%
	\begin{figure}[t]
	\label{fig:7} 
	\epsfclipon 	
	\centerline{\epsfig{file=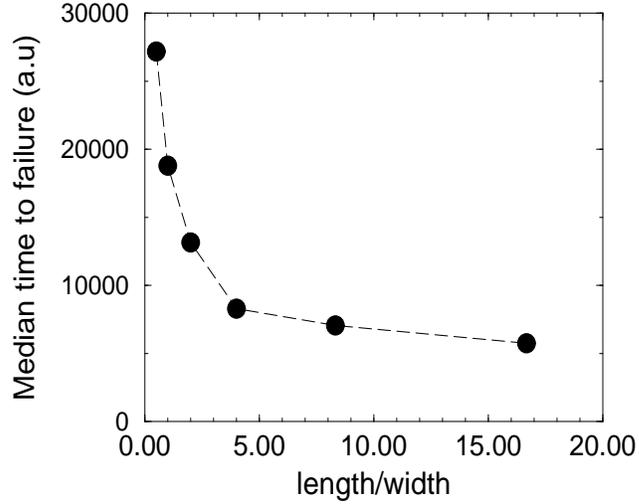,width=0.6
	\linewidth,height=0.5\linewidth}} 
	\caption{\protect\small Median time
	to failure as a function of the length to width ratio at fixed width.
	The network width and the parameter values are the same of Fig. 5.
	Points are the results of average over $20$ failure simulations.}  		\end{figure}
 		\begin{figure}[b]
		\label{fig:8} 
                \epsfclipon
                \centerline{\epsfig{file=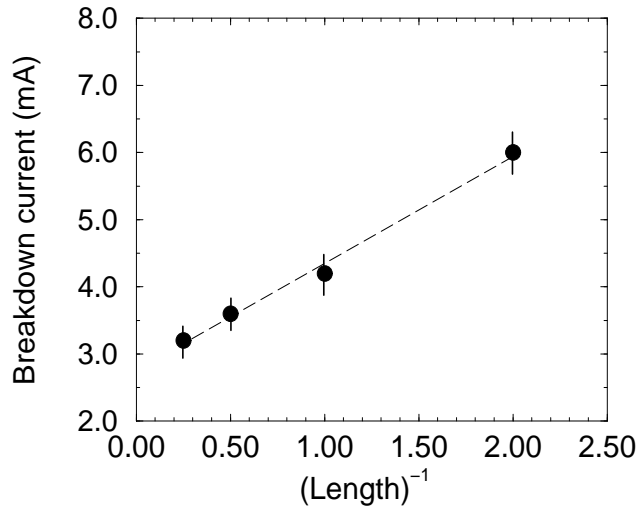,width=0.6
                \linewidth,height=0.5\linewidth}} 
		\caption{\protect\small Breakdown
                current as a function of the width to length ratio at
                fixed width. The network width and the parameter values are 
                the same of Fig. 5. The curve is a linear fit to data.} 			\end{figure}

\section{Conclusions}
We have developed a Monte Carlo approach to electrical breakdown phenomena, 
with particular attention on its applications to the study of electromigration
in metallic lines. Simulations are performed on a RRN whose configuration 
is evolving because of two competing percolations driven by an external 
current. The results of the simulations well agree with the experimental 
findings  concerning several features of the electromigration phenomenon, 
including compositional and geometrical effects. The flexibility of 
the approach allows wide possibilities of further extentions of the model. 
Future goals are the extraction of  information about precursor effects 
of the electical breakdown phenomena.
\end{document}